\documentclass[fleqn,10pt]{wlscirep}
\usepackage[utf8]{inputenc}
\usepackage[T1]{fontenc}
\usepackage{xcolor}
\usepackage{amsmath,amssymb,graphicx,epstopdf,bm,bbm}
\newcommand{\note}[1]{{\color{black} #1}}
\newcommand{\discuss}[1]{{\color{black} #1}}
\title{Augmenting knee biomechanics through programmable knitted ExoSkin orthoses}

\author[1,2,+*]{Krishma Singal}
\author[3,+]{Samuel P.~Kirschner}
\author[3,4,*]{Andrew K. Schulz}
\author[3]{Emily D. Sanders}
\author[3,5]{Gregory Sawicki}
\author[3]{Kinsey R. Herrin}
\author[1,6,*]{Elisabetta A. Matsumoto}
\affil[1]{School of Physics, Georgia Institute of Technology, Atlanta, Georgia 30332, USA}
\affil[2]{Department of Mechanical Engineering, Rice University, Houston, Texas 77005, USA}
\affil[3]{School of Mechanical Engineering, Georgia Institute of Technology, Atlanta, Georgia 30332, USA}
\affil[4]{Haptic Intelligence Department, Max Planck Institute for Intelligent Systems, Stuttgart, Germany}
\affil[5]{School of Biological Sciences, Georgia Institute of Technology, Atlanta, Georgia 30332, USA}
\affil[6]{International Institute for Sustainability with Knotted Chiral Meta Matter (SKCM$^2$), Hiroshima University, Higashihiroshima, Hiroshima 739-8526, Japan}

\affil[*]{ks251@rice.edu, sabetta@gatech.edu, aschulz@is.mpg.de}

\affil[+]{these authors contributed equally to this work}

\begin{abstract} 
A large subset of the population suffers injury or disease that causes knee pain and difficulty navigating day-to-day tasks. Off-the-shelf knee orthoses that are commonly used to treat these ailments overlook user-specific joint geometry and/or biomechanical needs. They can often be made with materials that lead to discomfort. We explore how the rich programmability of knitted fabrics can be harnessed to augment human biomechanics while promoting comfort. In this pursuit, we define \textit{ExoSkins}, a class of unpowered exoskeletons that are lightweight, comfortable, garment-like devices, and are designed based on user- and joint-specific needs. Although we foresee an expansive space for ExoSkin design (e.g., containing active materials, with sensing capabilities), in this study we focus on the interplay between knit geometry and programmable elasticity in passive orthoses as a means to augment the knee’s rotational stiffness. We design geometrically-programmed ExoSkins, abbreviated \emph{G-PExos}, that capitalize on the anisotropies of four types of knitted fabric to provide high stiffness for joint torque without the need for rigid materials. Our findings indicate that G-PExos can achieve rotational stiffness of similar magnitude to off-the-shelf orthoses and can also be tuned to achieve a much broader range of rotational stiffness without sacrificing comfort to the user.

\end{abstract}
\begin{document}

\flushbottom
\maketitle

\thispagestyle{empty}

\section*{Introduction}

\subsection*{Knee pathologies and orthoses}
The human knee joint plays a critical role in human activities ranging from general mobility to high intensity sports~\cite{markstrom2019dynamic}.
This complex joint houses many tendons, cartilages, bones, and muscles that all join together, forming a hinge with articulations that are susceptible to both acute and chronic injuries\cite{kumar_j_healthy_2020,abulhasan_anatomy_2017}. 
Ligamentous injuries of the knee are common with nearly 200,000 injuries to the anterior cruciate ligament (ACL) occurring each year in the US\cite{paterno_incidence_2014, frank_current_1997}. Once an ACL injury occurs, follow-up injuries are almost six times more likely to occur than in other healthy adults\cite{paterno_incidence_2014}.
Meniscal tears\cite{mordecai_treatment_2014, karia_current_2019}, which are acquired through trauma or natural degenerative processes, are another common condition experienced in the knee joint. Such tears can cause pain and/or major disruptions to the knee's biomechanics over time\cite{mordecai_treatment_2014}. 
One of the most prominent long-term knee pathologies, knee osteoarthritis (KOA), is associated with diminished articular cartilage in the joints and results in increased bone-to-bone friction, general pain and discomfort, and a decrease in quality of life due to reduced capacity for standing, walking, and climbing stairs\cite{felson_prevalence_1987,murphy_lifetime_2008,budarick_design_2020,mandl_osteoarthritis_2019}. KOA impacts a significant portion of the population with some evidence suggesting that the lifetime risk of symptomatic KOA is as high as 45\%\cite{murphy_lifetime_2008}. Additionally, these pathologies are related as there is a nearly 50$\%$ chance of developing KOA within 10 years of undergoing ACL surgery\cite{webster_anterior_2022}. 
Since the knee joint is crucial to an individual’s mobility and independence, non-invasive treatment modalities to delay, avoid, or aid recovery in surgical intervention, restore function, or reduce pain are needed~\cite{derogatis2019non}.

Knee braces and sleeves (i.e., knee orthoses) are the most common intervention used to prevent injuries, promote recovery, or relieve pain. Four main types of passive knee orthoses are currently available on the market: 1) \emph{prophylactic knee orthoses} are intended to prevent injury by restricting motion during athletics; 2) \emph{functional knee orthoses} are intended to decrease the risk of reinjury by providing stability during athletics; 3) \emph{rehabilitative knee orthoses} are intended to promote rehabilitative success by limiting joint motion and providing protection; and 4) \emph{patellofemoral knee orthoses} are intended to promote proper patellar movement that can relieve pain behind the kneecap\cite{paluska_knee_2000,dzidotor_functions_2024}. New categories of passive orthoses are being defined in research and development, where attempts have been made to displace body weight from damaged knees using pneumatic orthoses\cite{stamenovic_pneumatic_2009} and 3D printed components\cite{elliot_customized_2021}. Quasi-passive and active orthoses constitute a dynamic field of research\cite{stoltze_development_2022,fang_novel_2020} and have recently emerged as commercially available devices\cite{kentin_flexible_2024}, though all depend on exogenous energy sources. 

Despite having a global market worth \$1.19 billion in 2024\cite{grand_view_research_gvr_2024}, knee orthoses have had limited success in achieving desired user outcomes. In off-the-shelf orthoses, elastomeric materials, like neoprene, and knitted fabrics are embedded with elastomeric fibers are typically used in a \emph{one-size-fits-most} strategy, since elastomeric materials can stretch over a wide range of human anatomies\cite{pereira_study_2007}. They work by applying compression once they have been donned\cite{singha_analysis_2012}. Not only do these off-the-shelf orthoses lack a user-specific design that may cause the device to be ineffective and/or uncomfortable due to poor fit, but popular materials, such as neoprene, also suffer from challenges associated with breathability and comfort against the skin during prolonged wear (e.g., chafing or blistering)\cite{pereira_study_2007}. 
Cost, lack of comfort, bulkiness, and weight are additional prominent factors leading users to discontinue the use of knee orthoses\cite{beaudreuil_clinical_2009, dessery_comparison_2014, shamaei_biomechanical_2015}. Even when used correctly and consistently, there is ongoing debate surrounding the effectiveness of these orthotic devices at preventing injury and promoting long-term pain relief. For example, past studies investigating the effects of knee orthoses on treating KOA show little lasting impact on an individual’s prognosis\cite{beaudreuil_clinical_2009}. While orthoses aimed at unloading forces on the knee joint have been shown to diminish pain, they do not demonstrate definitive long-term benefits in relieving pain as they are not consistently prescribed and often result in discomfort\cite{beaudreuil_clinical_2009, kirkley_effect_1999}. Additionally, while there are specific sports (e.g., skiing) where knee orthoses have been shown to reduce post-operative injuries\cite{smith_functional_2014}, there remain questions about the functional benefits of such orthoses across other sports. 

Currently available orthoses fail to factor more accurate models of knee function and geometry into their design, missing the opportunity to integrate advancements in the understanding of biomechanics. The human knee can be modeled as a one degree-of-freedom (DoF) hinged joint with phases of flexion and extension ranging approximately 0-150\textdegree. During a walking cycle, one can define specific phases: heel contact, maximum flexion, end of extension, terminal stance, pre-swing, and swing\cite{rose_walking_1994}. Understanding the spring-like properties of the knee is necessary for designing an orthosis that optimizes the knee's stiffness to assist motion. The stiffness of the knee ranging from flexion to extension to weight-acceptance is around 260-300 Nm/rad\cite{shamaei_estimation_2013}. The knee stiffness during this range is what needs most support, as this is when the lower limb must support the body weight of the individual. Recognizing the shortcomings of available orthoses and leveraging advancements in biomechanics, we are motivated to pursue an improved strategy for designing knee orthoses.

\subsection*{ExoSkins}
We define \textit{ExoSkins} to be a new class of soft unpowered exoskeletons that are lightweight, garment-like, wearable devices designed based on user specific ailments and anatomies.
Within the ExoSkin framework, we work to move away from the one-size-fits-most mindset by developing a personalizable framework to create orthoses that are more effective and comfortable than current off-the-shelf options. 
There has been development of exoskeletons and exosuits which help offset loads from the back, shoulder, or arms\cite{zhou_portable_2024,chung_lightweight_2024,arens_preference-based_2025,arnold_personalized_2025,lamers_feasibility_2018}; however, these innovations utilize motors, soft pneumatics, or cables.
In this paper, we specifically focus on ExoSkins made only from knitted fabrics and investigate their ability to achieve desired mechanical function. We fabricate knitted knee orthoses via strategic patterning of knitted stitches that form regions with different anisotropic elastic properties\cite{singal_programming_2024}, all while only using a nearly inextensible yarn.
Knitting is an additive manufacturing technique that, with the help of industrial knitting machines, can form seamless fabrics encoded with bespoke mechanical properties and embedded with high performance materials. 
Knitted fabrics have already been used as spacer fabrics in orthotic devices\cite{yu_advanced_2025} to provide a more comfortable interface with human skin, but there is also evidence suggesting their potential to provide support and increased range of motion in orthoses that make use of knitted fabrics as the main structural component\cite{legner_medical_2001, pereira_study_2007, singal_programming_2024}.

A common example of knitting in medical applications is compression stockings that improve circulation and reduce edema\cite{legner_medical_2001,lim_graduated_2014}. These stockings are normally designed for maximum compressive forces at the ankle that gradually decrease towards the groin and can reach up to 50 mm Hg on parts of the leg\cite{lozo_structure_2022}. Compression stockings primarily use two yarns: the first yarn is knit into a traditional stockinette fabric (also known as jersey fabric) and the second is woven through the loops in a ``bent weft" fashion (cf. Fig. 3 in Lozo et al.\cite{lozo_structure_2022}).
The knitting stitch pattern takes advantage of the inlaid elastomeric yarn to provide circumferential compression but the one-size-fits-most strategy does not take advantage of the geometric customizability afforded by knitting.
Compressive garments typically employ elastic yarns and include \emph{negative ease}, where the garment dimensions are smaller than the anatomic measurements, to increase the elastic modulus circumferentially around the leg and apply variable compressive forces along the leg. Other knitted compression garments have been fabricated with embedded shape-memory alloys that allow for thermally-driven actuation\cite{granberry_functionally_2019, granberry_dynamic_2022}.

\begin{figure}
    \centering
    \includegraphics[width = 1\textwidth]{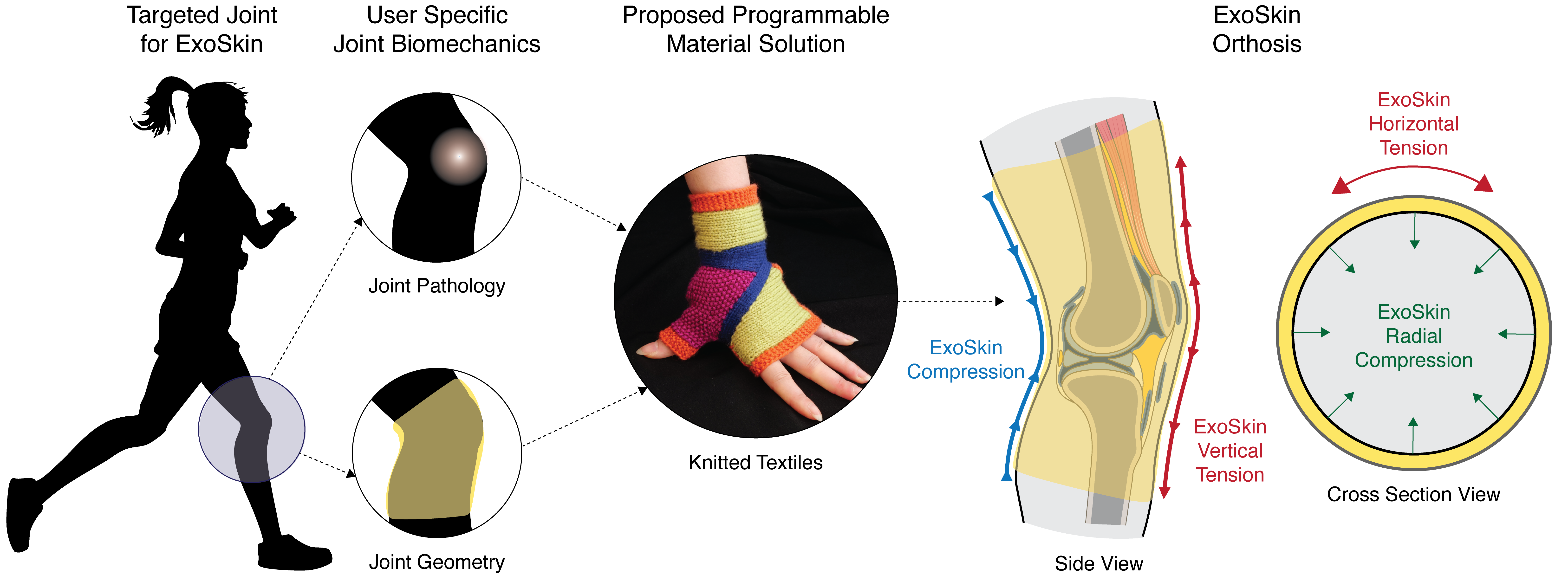}
    \caption{
   \textbf{Design methodology for ExoSkin orthoses.} After targeting a specific joint, user biomechanics and morphology are considered in designing the knitted ExoSkin with an informed intuitive based approach. Previous work by Singal et al.\cite{singal_programming_2024} designed a therapeutic glove prototype informed by mechanical characterization of different types of fabric. We integrate this approach to design an orthosis for the knee. Shear forces (primarily in tension) acting on the exterior of the knee influence the biomechanics.
    }
    \label{fig:fig1}
\end{figure}

We hypothesize that by strategically programming spatially varying anisotropies of knitted fabrics into passive ExoSkins (Figure \ref{fig:fig1}), we can create a programmable knee orthosis, specific to the geometry and biomechanical needs of the wearer, that is as effective, if not more so, than comparable, commercially available, off-the-shelf orthoses.
We aim for these geometrically-programmed ExoSkins, \emph{G-PExos}, to conform not only to the user's geometry but to be specifically tailored with torque-angle responses in the same range of response as off-the-shelf orthoses, taking advantage of the customizability of knitted fabrics\cite{choi_effect_2000}. 
We first characterize and compare the mechanical properties of orthoses composed of common knitted stitch patterns (simple knitted ExoSkins) and several existing off-the-shelf orthoses for the knee. Swatches of material taken from the orthoses are mechanically characterized via uniaxial stretching experiments. We follow a similar approach that Singal et al.\cite{singal_programming_2024} used in designing a knitted therapeutic glove prototype to design and fabricate G-PExos with tailored spatial variations in knitted stitches that take advantage of the unique anisotropic elastic properties of each knitted stitch pattern. Finally, we compare the torque-angle response of the G-PExos to the off-the-shelf knee orthoses on a custom test rig intended to mimic human knee flexion.
Successful development and deployment of these G-PExos will allow us to further expand this intuitive design process to design ExoSkins for a diverse array of joints across the human body. For individuals with various joint pathologies (e.g., KOA, pain, or ligamentous tears or instabilities\cite{gianotti_incidence_2009,heidari_knee_2011}), ExoSkins could improve quality of life by alleviating pain and assisting with post-treatment healing. 
For healthy individuals, these knitted ExoSkins may be able to provide supraphysiological augmentation of joints and limbs\cite{collins_reducing_2015}.

\section*{Results}

\subsection*{Simple knitted ExoSkin construction and comparisons with off-the-shelf orthoses} 
To analyze the current off-the-shelf orthosis performance, we acquired three off-the-shelf orthoses (Figure \ref{fig:fig2}a) to test: a ``neoprene orthosis" (McDavid), an ``elastic orthosis" (Mueller Sport OmniForce Knee Support K-100 - SM brace), and a ``rigid orthosis" (McDavid 425 Ligament Knee Support with Stays and Cross Straps). These three orthoses were selected to represent the dominant trends observed in commercially available off-the-shelf knee orthoses. 
The neoprene orthosis is made of neoprene, a synthetic rubber, which is a fundamentally different material from our knitted fabrics.
The rigid orthosis contains two embedded chains of rigid links adjacent to the kneecap and velcro straps that can be used to adjust radial compression to the leg (\note{see Supplemental Figure S1}).
The elastic orthosis is closest to our custom knitted ExoSkins, but its knitted fabric includes elastomeric yarn interwoven into the knitted stitches (see Figure \ref{fig:fig2}a inset). The fabric is double knit such that each face of the fabric is composed of a separate yarn with another elastic yarn inlaid between.

We compared these off-the-shelf orthoses to our custom fabricated knitted orthoses. The process of knitting involves manipulating yarn into a rectangular lattice of slipknots known as stitches, the two most basic being the knit stitch and the purl stitch, shown in black and gray, respectively, in Figure \ref{fig:fig2}b. How one patterns these two types of stitches impacts the final knitted fabric structure and its elastic properties. We considered four types of knitted fabric in this study: stockinette is made from only knit stitches, garter is made from alternating rows of knit and purl stitches, rib has alternating columns of knit and purl stitches, and seed is a checkerboard-like lattice of the two stitches (see Figure \ref{fig:fig2}b). Using an industrial knitting machine, we constructed three ``simple knitted ExoSkins'' made of a single stitch pattern (stockinette, garter, and seed) to mechanically test alongside the off-the-shelf orthoses (see \note{Supplementary Figure S2}). We did not make a simple knitted ExoSkin composed of the rib stitch pattern. The relative sizes of the simple knitted ExoSkins are comparable to the off-the-shelf orthoses when relaxed.

\subsection*{Swatch-level mechanical properties}

\begin{figure}
    \centering
    \includegraphics[width = 1\textwidth]{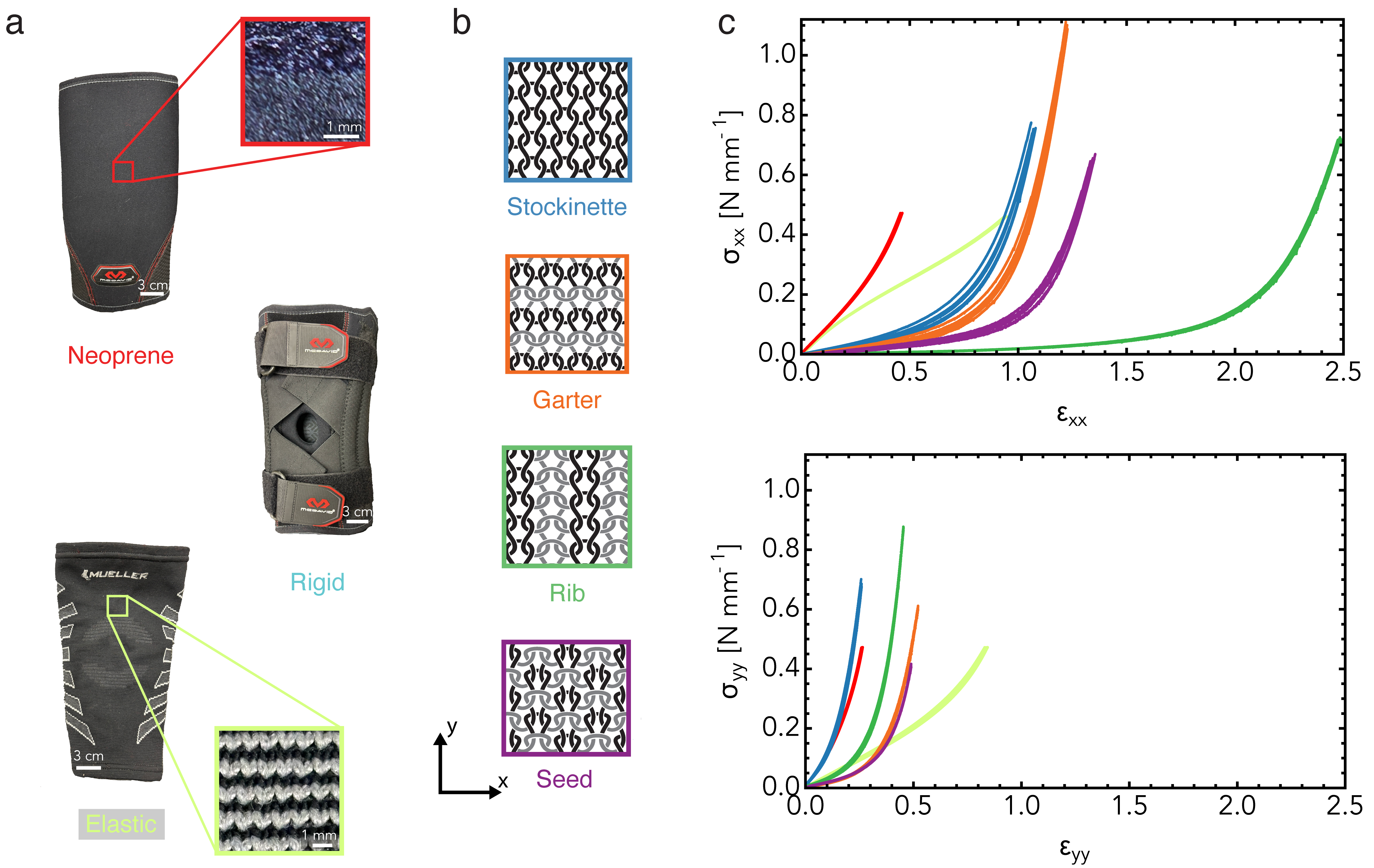}
    \caption{
    \textbf{Uniaxial tension tests of fabric swatches taken from off-the-shelf and simple knitted ExoSkins.} We compare swatches from the off-the-shelf orthoses shown in (\textbf{a}) with swatches of stockinette, garter, rib, and seed knitted fabrics made with acrylic yarn. The stitch patterns for the four types of knitted fabrics are shown in (\textbf{b}), where black is a knit stitch and gray is a purl stitch. The stress versus strain response of each swatch is shown in (\textbf{c}), where the colors correspond to: stockinette (blue), garter (orange), rib (green), seed (purple), neoprene (red), and elastic (lime-green).
    }
    \label{fig:fig2}
\end{figure}

We performed uniaxial stretching experiments on swatches of fabric from two of the three off-the-shelf orthoses and four types of knitted fabrics (see Methods and \note{Supplementary Figure S3}). We did not consider a swatch from the rigid orthosis since there are numerous components that add to its support (\note{see Supplementary Figure S1}), and there is no clear way to select a swatch that captures its material make-up. The 2D stress, $\bm{\sigma}$, versus strain, $\bm{\epsilon}$, relationship for each swatch is shown in Figure \ref{fig:fig2}c. The stress versus strain relationship along the knitted fabrics' $x$- and $y$- directions (refer to the coordinate frame in Figure \ref{fig:fig2}) showcases the anisotropies and distinct strain-stiffening behaviors of the different types of knitted fabrics, in contrast to the materials comprising the off-the-shelf orthoses, which exhibit much less drastic strain-stiffening and have much closer properties in the two directions explored here. In the knitted fabrics, low stiffness was observed initially, while the stitches were free to deform locally. Once the fabric became taut, the yarn itself was put into tension and the mechanical response of the fabric captured the interplay between the stitch geometry and elastic properties of the constituent yarn. 

We extracted the Young's moduli (tensile stiffness) in the $x$- and $y$-directions (i.e., $Y_x$ and $Y_y$) of each swatch as the slope of its stress-strain curve in the low strain (linear) region (see Methods and \note{Supplementary Figure S4}). The $x$-direction stiffness contributes directly to the ease with which the ExoSkin can be stretched over the knee and correlates to the magnitude of radial stress (compression) acting on the leg once the orthosis is in place. The $y$-direction stiffness provides the dominant resisting force from the orthosis as the knee rotates. 

In the $x$-direction, we found that the swatches from the off-the-shelf orthoses are over four times stiffer than the stockinette and garter knitted fabrics and an order of magnitude stiffer than the rib and seed knitted fabrics (Figure \ref{fig:fig2}c, Table \ref{tab:rigidity2}). The high $x$-direction stiffness observed in the off-the-shelf materials is likely needed to accommodate the one-size-fits-most design strategy often employed for commercial clothing\cite{christel_average_2017}, healthcare\cite{mcandrews_one_2022}, and electronics\cite{straczkiewicz_one-size-fits-most_2023}. It is also noted that in the $x$-direction, the elastic swatch exhibits slight strain softening, while all other swatches exhibit strain stiffening, although strain stiffening in the knitted fabrics is much more apparent than in the neoprene swatch. Additionally, while all of the knitted fabrics are softer than the off-the-shelf swatches in the $x$-direction, the rib knitted fabric is noticeably the softest and reaches nearly 200\%  of its resting length before strain-stiffening behavior is exhibited. The neoprene swatch is the stiffest in the $y$-direction (Figure \ref{fig:fig2}c, Table \ref{tab:rigidity2}) followed by stockinette, which exhibits a six fold increase in stiffness relative to its $x$-direction Young's modulus. Rib and seed similarly have higher stiffness in the $y$-direction than in the $x$-direction, with rib showing a sixfold increase and seed almost doubled. Garter, on the other hand, has similar stiffness in both directions. It is interesting to note that the materials from the off-the-shelf orthoses have a similar range of response as the custom knitted fabrics in the $y$-direction but are noticeably stiffer in the $x$-direction.

These swatch-level experiments indicate that ExoSkins composed of our custom knitted fabrics may be easier to put on and exhibit similar rotational resistance as the off-the-shelf orthoses, but will impose smaller radial compression as long as the knitted fabrics are not taut during use. However, uniaxial stretching experiments are not enough to fully characterize the mechanics of the orthoses since more complex stress states are expected when the fabrics conform to the three-dimensional knee. Additionally, although the knee joint has one primary degree of motion, the compressive nature of the orthoses requires us to consider anisotropy of the material used throughout the orthosis. In the next section, we detail an experimental protocol that considers the three-dimensional aspects of the knee and how the the swatch-level mechanics relates to on-knee use.

\subsection*{Torque-angle measurements of orthoses and simple knitted ExoSkins undergoing knee flexion}

We experimentally characterized the torque-angle response of the simple knitted ExoSkins and the off-the-shelf orthoses using an in-house test rig designed to mimic human knee flexion (Figure \ref{fig:fig3}a-b and see Methods). 
The torque versus angle relationships for each of the tested orthoses are shown in Figure \ref{fig:fig3}c. The simple knitted ExoSkins all contained a linear regime followed by strain-stiffening behavior, similar to the individual swatch behavior observed under uniaxial tension. Each of the off-the-shelf orthoses were close to linear over the entire joint rotation, with some minor strain softening at large angles. It is important to consider the entire range of motion in comparing the different orthoses. 
For small rotation angles ($\theta \leq 20^\circ$), the neoprene and elastic orthoses had similar rotational stiffness falling in the mid-range of the orthoses tested. 
The rigid orthosis and the stockinette simple knitted ExoSkin had much higher rotational stiffness, while the garter and seed simple knitted ExoSkins had lower rotational stiffness than any other orthosis in the small angle range. Because of the varying degrees of strain stiffening behavior exhibited by the simple knitted ExoSkins, they covered a much wider range of rotational stiffness as angle increases as shown in Figure \ref{fig:fig3}d, where the rotational stiffness at $\theta = 30^{\circ}$ is compared for the different orthoses. 
In Figure \ref{fig:fig3}c, we also compared the torque-angle responses to published values of human knee rotational stiffness (blue dashed line) and 5$\%$ of the human knee rotational stiffness\cite{shamaei_biomechanical_2015} (forest green dashed line), and note that none of the orthoses investigated here reached 5$\%$ of the human knee rotational stiffness.

We compared the swatch-level uniaxial stretching results with the orthosis-level torque angle experiment results in \note{Supplementary Fig. S5}. The swatch-level stress-strain data and the orthosis-level torque-angle data were converted to force-angle data for more direct comparison. Neither the $x$-direction (\note{Supplementary Fig. S5}a) nor the $y$-direction (\note{Supplementary Fig. S5}b) swatch-level data matched the orthosis-level data (\note{Supplementary Fig. S5}c), which indicates that stress-states beyond uniaxial tension play an important role in dictating the mechanical response of knee orthoses and experimental protocols designed to mimic knee motion are critical.

\subsection*{Fabrication of geometrically-programmed ExoSkins (G-PExo)}

The variability of mechanical response observed in the knitted fabric swatches and simple knitted ExoSkins inspired us to explore the potential of mixing different knitted fabrics in a single ExoSkin in order to tune the rotational stiffness and promote comfort.
We constructed two knitted ExoSkins made of a combination of the four types of knitted fabrics (stockinette, garter, rib, and seed). 
These geometrically-programmed ExoSkins (G-PExos) capitalize on the anisotropies of the four knitted fabrics (recall Figure \ref{fig:fig2}c). 

The front of both G-PExo1 and G-PExo2 are primarily designed to capitalize on the $y$-direction properties of the knitted fabrics, while also keeping $x$-direction properties in mind (see Figure \ref{fig:fig2} and \note{Supplementary Figure S6}). We used the stiffest knitted fabric, stockinette (see Table \ref{tab:rigidity2}), in two curved stripes extending from the upper to lower limb. By avoiding the knee cap, these stripes provide high rotational stiffness without impeding motion. Stockinette is the primary load carrying fabric, whereas the remainder of the design is meant for comfort and fit.
Rib is used for the cuffed portions on the top and bottom since it is relatively stiff in the $y$-direction but its extensibility along the $x$-direction facilitates stretching the ExoSkin over the leg.  Seed, being one of the softest fabrics along both its axes, is placed on the kneecap for maximum extensibility and to provide flexibility and comfort. The user can comfortably flex their knee without additional stress on the kneecap. Garter is used for the remainder of the front panel to allow the orthosis to rest flat on the leg while allowing for some extensibility in the $y$-direction. Garter is especially stretchy along its $x$-direction so the garment still comfortably stretches around the circumference of the leg. 
The front panels of G-PExo1 and G-PExo2 are identical except that the stockinette stripes of G-PExo2 are about 2.2 times wider (\note{Supplementary Figure S6}) than those of G-PExo1 (about 0.8 cm for G-PExo1 and 1.75 cm for G-PExo2). As a result, we expected G-PExo2 to provide higher rotational stiffness.

The back of both G-PExos (Figure \ref{fig:wearingsleeve}b) were designed to capitalize on the $x$-direction properties of garter and rib acting in the $y$-direction of the G-PExo (see Figure \ref{fig:fig2}). Because flatbed machine knitting involves building fabrics along their $y$-axis, we constructed the back of the G-PExos on the machine and then rotated them by $90^{\circ}$ before attaching them to the front half of the G-PExo. Garter made up a majority of the back panel to provide flexibility and make it easy to slip on. Rib fabric was used directly behind the knee, which enables it to stretch easily when the knee is straight and relax when the knee is bent. This back panel design was intended to improve wearability, comfort, and fit by mitigating bunching behind the knee and providing a snug fit without adding unnecessary compressive forces on the leg. Once fabricated on the knitting machine, the front and back panels were sewn together, as shown in \note{Supplementary Figure S6}, and then donned (Figure \ref{fig:wearingsleeve}c).

\subsection*{G-PExo characterization and comparisons}

Similarly to the off-the-shelf orthoses and the simple knitted ExoSkins, we tested our G-PExos on our in-house test rig mimicking human knee flexion and extracted the torque-angle curves (Figure \ref{fig:fig3}c).
G-PExo1 falls between the stiff stockinette ExoSkin and the more flexible garter and seed ExoSkins. G-PExo2, with thicker stockinette stripes, is significantly stiffer than G-PExo1 and has a torque angle response that almost exactly matches the stockinette ExoSkin. Just by increasing the thickness of the stockinette stripes of G-PExo2, we create a knitted orthosis that has higher rotational resistance.

Compared to other knitted compression garments, our G-PExos enable tunable control of rotational stiffness without sacrificing comfort by making the entire fabric tighter or using elastomeric yarn. We maintain the same negative ease and geometric shape between both G-PExos. We have previously highlighted the anisotropic properties of the different types of knitted fabrics and the varying biaxial strain resulting from their use. Thus, the circumferential tension in the orthosis is impacted from how the knit fabrics are utilized. By changing where the different types of knitted fabric are geometrically arranged, we decouple the geometric size of the orthosis from the tension in the fabric at prescribed locations of the orthosis. 

Both G-PExos are stiffer than the neoprene and elastic orthoses, and the tunability of rotational stiffness demonstrated in the G-PExos indicates the versatility of knitted fabrics for user-specific orthoses. \discuss{The rigid orthosis does possess geometric patterning and components designed to increase stiffness (\note{Supplementary Figure S1}), but the G-PExos possess better dynamic control.} We also note that all of the off-the-shelf orthoses are linear or strain-softening, whereas all of the knitted ExoSkins are strain-stiffening. By varying stitch patterns, and consequently concentrations of different types of knitted fabrics, the knitted orthoses can potentially provide a delayed stiffness effect for specific angles. \discuss{This could encourage users to not overextend joints and prevent injuries.}

To systematically compare each orthosis, we extracted the stiffness at $\theta = 30^{\circ}$ (i.e., the slope of the curve fit at that point) and compared the off-the-shelf orthoses, simple knitted ExoSkins, and G-PExos (Figure \ref{fig:fig3}d). At $\theta=30^{\circ}$, all three types of orthoses have very low stiffness, with the least stiff off-the-shelf orthosis (elastic) having only one-tenth the stiffness of the human knee (Figure \ref{fig:fig3}d). The rigid orthosis is the stiffest of off-the-shelf orthoses but only reaches around $2\%$ of human knee stiffness. The knitted ExoSkins have a much more extensive range. The stockinette ExoSkin reaches roughly $4.5\%$ of human knee stiffness while the garter and seed orthoses exhibit stiffnesses on par with the off-the-shelf orthoses (Figure \ref{fig:fig3}d). 
G-PExo1 has a significantly increased stiffness compared to all of the off-the-shelf orthoses and the seed and garter ExoSkins (roughly $4\%$ of the human knee) with only $8\%$ of the front panel making use of stockinette, the stiffest of the knitted fabrics. 
G-PExo2, which more than doubles the stockinette stripe thickness of G-PExo1, achieves an equivalent stiffness value to that of the stockinette ExoSkin despite only having $20\%$ of the front panel composed of the stockinette fabric. G-PExo2 maintains a high percentage of garter and seed for comfort and fit, yet is the closest of the orthoses in this study to $5\%$ of human knee stiffness. 
These results highlight the programmable potential of knitted fabrics in orthosis design and construction (Figure \ref{fig:fig3}c-d). 

\section*{Discussion}

In this work, we introduced a class of soft unpowered exoskeletons, i.e., ExoSkins, which are user-specific, lightweight, form-fitting, and programmable wearable devices aimed to assist or augment human biomechanics. We specifically focused on ExoSkins for the knee joint that could impact knee flexion and sought to obtain a torque-angle response comparable to off-the-shelf orthoses with additional tunability to adapt to user specific needs. By strategically mapping different knitted fabrics, based on mechanical characterization of fabric swatches, we designed geometrically-programmed ExoSkins, G-PExos, and compared to off-the-shelf orthoses.

We conclude that our G-PExos possess torque-angle responses comparable to off-the-shelf knee orthoses. Furthermore, we can tune the torque-angle response of the G-PExos by tuning its design without making the entire fabric tighter or sacrificing comfort. 
Our G-PExos reach close to $5\%$ of human knee stiffness, which indicates their viability to support the knee joint and affect its response. These results indicate a promising avenue toward tailoring the torque-angle behavior of knitted ExoSkins by altering their spatially-varying stitch patterns and the yarns used to fabricate them. The neoprene and elastic off-the-shelf orthoses compress uniformly on the leg while the G-PExos have tailored regions of directed strain and comfort. Even with our balance of function and comfort in the design of the G-PExos, we are still within the realm of response of common off-the-shelf orthoses. 
The knitted orthoses, whether composed of a single or multiple types of fabric, are breathable, soft, and comfortable, while maintaining functionality.    

This study demonstrates that our custom ExoSkins have superior tunability in rotational stiffness, yet it is still apparent that off-the-shelf orthoses are designed to also provide uniform compression to the leg, a quantity that was not studied here due to the limitations of our rigid test rig. We have so far taken into account a specific individual's geometry and joint needs to design tailored ExoSkins for the knee (and the wrist \cite{singal_programming_2024}) (see Figure \ref{fig:fig5}a), but further user testing will better help assess its effectiveness. Additional investigation (likely as a clinical trial) is needed to understand the relevance of using ExoSkins for augmenting rotational stiffness, joint compression, energy expenditure, or providing other application-specific biomechanical augmentations. This will better inform how ExoSkins compare to state-of-the-art exoskeleton devices for the knee\cite{lee_mitigating_2024,lee_reducekneehyper_2023}. Further investigation is also needed to understand whether the ExoSkins will suffer from stress-relaxation that could compromise their performance over time. Former studies have found that stress relaxation is higher in the $x$-direction than the $y$-direction\cite{saufun_ng_yip_medical_1994} and others have reported that the use of tuck stitches can decrease the stress relaxation along the $y$-direction while increasing it along the $x$-direction\cite{asayesh_analysis_2024}.

Design optimization of the spatial distribution of stitches and/or stitch patterns, which has recently been explored in upper limb prostheses\cite{pasquier_knit_2025}, could further expand the range of rotational stiffness or other properties achieved in our custom knitted ExoSkins. Such an optimization-based approach would also promote user-specific biomechanical considerations (see Figure \ref{fig:fig5}a) during design (e.g., external dimensions, range of motion, muscle capacity). For example, we could measure an individual's unique biomechanics (e.g., with a dynamometer or motion capture system) and the difference between the user's biomechanics and an ideal model could be used as input for the optimization. 

To add additional programmability to our knitted ExoSkins beyond stitch geometry, future work could consider modulating the mechanical properties of the constituent yarn (e.g., from highly extensible yarns to very stiff yarns containing Kevlar or carbon fiber forming a composite yarn) or embed active yarns that provide actuation capabilities\cite{wu_biomimetic_2023, granberry_functionally_2019, granberry_dynamic_2022}. Refinement of the attachment mechanism (e.g., cable-ratchet tensioners, harness system) will also be needed to prevent slipping of the ExoSkin against the body to ensure efficient transfer of forces while maintaining user comfort. Furthermore, by incorporating sensing-feedback systems into our ExoSkins, we could take advantage of specialized materials that respond to mechanical\cite{jumet_fluidically_2023}, electrical\cite{yang_toward_2021}, or thermal stimuli\cite{weinberg_multifunctional_2020}. We will also pursue improvements to ExoSkin fabrication. We are working on a seamless knee orthosis design that does not require sewing. The limitation to seamless designs, at least when fabricated on flatbed industrial knitting machines, is that they must use the $y$-direction properties of the knitted fabrics. However, a seamless design would reduce the manual labor required to assemble the G-PExos studied here. 

Nevertheless, this study highlights the potential for supporting and enhancing biomechanical performance using spatially-varying stitch patterning alone, and other use-cases can be explored as future work such as for the elbow and shoulders (see Figure \ref{fig:fig5}b).
In the future, ExoSkins, and further tuned G-PExos, could improve the agility, stability, and economy of natural movement with the advantage of being entirely passive devices that are tunable to a user's specific biomechanics and joint anatomy.

\section*{Methods}

\subsection*{Fabrication Tools}

The knitted orthoses, both the simple ones and the G-PExos, were designed with STOLL's M1 Plus software and subsequently fabricated on a STOLL CMS 530 HP Industrial Knitting Machine. The simple knitted orthoses were constructed using the Tamm Petit 2/28 T4201 Red 100\% acrylic yarn from The Knit Knack Shop\texttrademark{}. The G-PExos were constructed using the Tamm Petit 2/30 T4201 White 100\% acrylic yarn from The Knit Knack Shop\texttrademark{}.

The general shape and construction of the custom ExoSkins mimic that of a basic off-the-shelf knee orthosis. They consist of two flat pieces of fabric (one for the front of the leg and one for the back of the leg) with dimensions based on one of the author's leg measurements (18-inch thigh circumference, 15-inch shank circumference, and 12-inch ExoSkin height). To ensure a close fit, we consider negative ease in the design by fabricating each piece's top and bottom edge 1-inch smaller than the actual thigh and shank half-circumferences (i.e., 8-inches proximally, 6.5-inches distally, and linearly interpolated between these two values along the 12-inch ExoSkin height). When the front and back pieces are sewn together along their tapered edges, the resulting ExoSkin assumes a conical frustum shape (\note{see Supplementary Figure S6}). The pieces were sewn using a Bernette B77 sewing machine.

\subsection*{Swatch-level Characterization Experiments}

Swatch-level uniaxial stretching experiments were performed on an Instron Universal Testing Machine (UTM) Model 68SC-1 with a 500 N load cell (\note{Supplementary Figure S3}). We stretched swatches of the four types of knitted fabrics and two of the off-the-shelf orthoses along their $x$- and $y$-axes at a rate of 0.5 mm/s to 25 N for the knitted fabric swatches and 30 N for the off-the-shelf orthoses swatches (\note{Supplementary Figure S3}). One swatch of each orthosis underwent six trials of uniaxial tests. The first trial is taken out and treated as an initialization run meant to remove any handling bias. The swatches were clamped with custom 3D-printed clamps with small teeth to prevent the samples from slipping when being stretched (\note{Supplementary Figure S3b}). The off-the-shelf orthoses swatches were 2.5 by 2.5 inches (\note{Supplementary Figure S3c}). The knitted fabric swatch dimensions varied due to differences in their relaxation behavior, but each knitted swatch was made of 32 rows and 32 columns of knitted stitches (\note{Supplementary Figure S3d}).

We calculate the bulk stress, $\sigma = \frac{F}{W}$, and bulk strain, $\varepsilon = \frac{L-L_o}{L_o}$, of the samples while undergoing tension along the samples' $x$- and $y$- axes where $F$ is the applied force, $W$ is the width of the swatch at the top clamped end, $L_o$ is the initial unstretched length of the swatch, and $L$ is the stretched length of the swatch (see Supplementary Figure S3). Specifically, the $x$-direction stress and strain are defined as $\sigma_x = \frac{F_x}{W_y}$ and $\varepsilon_x = \frac{L_x-L_{x,o}}{L_{x,o}}$, respectively, and the $y$-direction stress and strain are defined as $\sigma_y = \frac{F_y}{W_x}$ and $\varepsilon_y = \frac{L_y-L_{y,o}}{L_{y,o}}$, respectively. 

We compared the Young's moduli of the off-the-shelf and knitted swatches in the $x$- and $y$-directions by performing linear least squares regression with linear fit and extracting the slope of the fit (\note{see Supplementary Figure S4}). In performing the regression for a given swatch, data from all five tensile experiments are considered together. We consider the linear 2D elastic regime to be $\epsilon_x \leq 0.50 \times Max[\epsilon_x]$ and $\epsilon_y \leq 0.25 \times Max[\epsilon_y]$ for the $x$- and $y$-directions, respectively.
With these linear fits, the Young's modulus is calculated as $Y_x = \frac{\sigma_{x}}{\varepsilon_{x}}$ and $Y_y = \frac{\sigma_{y}}{\varepsilon_{y}}$, where $Y_x$ is the $x$-direction Young's modulus and $Y_y$ is the $y$-direction Young's modulus. To evaluate the goodness of fit, we calculated the coefficient of determination, $R^2$, for each fit. The Young's moduli and $R^2$ values for all swatches are included in Table \ref{tab:rigidity2}. 

\subsection*{Test rig Construction}
A 26 in $\times$ 26 in $\times$ 18 in rigid frame consisting of 1 in$^{2} $ T-slot aluminum extrusion served as the structure to which the foam knee replica and pulley were mounted. To construct the knee replica, a plaster cast was taken of one of the author's legs (\note{see \note{Supplementary Figure S7}}). A two-part curable foam (Pedilen Rigid Foam 700) was poured into the cast, and a one-inch diameter PVC pipe was set into the center line during curing. Once the foam was set in the mold, the cast was removed and the replica was cut at the knee joint to separate it into upper and lower parts. Additional material at the posterior knee (popliteal region) was removed with a band saw to allow for a typical walking flexion of motion (approximately 70°) of the replica knee. To function as a joint, a hinge was inserted into the PVC pipe on both the upper and lower parts of the leg. The hinge was composed of two 3D-printed components made of PLA that interlocked with an aligned through-hole. A 1/4-inch bolt secured the two components, allowing them to pivot around the bolt. Once assembled, the proximal thigh was fastened to the test rig using a 3D printed fixture, and the distal shank was allowed to rotate freely. A 1/16 in braided steel cable with low strain under load was attached to the bottom of the lower leg using a 3D-printed insert that fit into the PVC pipe. The cable was then routed through a pulley and connected to a handheld force meter. By pulling downward on the force meter, rotation (flexion) was induced in the distal shank while measuring tension.

\subsection*{Torque angle experiments}

\begin{figure}
    \centering
    \includegraphics[width = 0.9\textwidth]{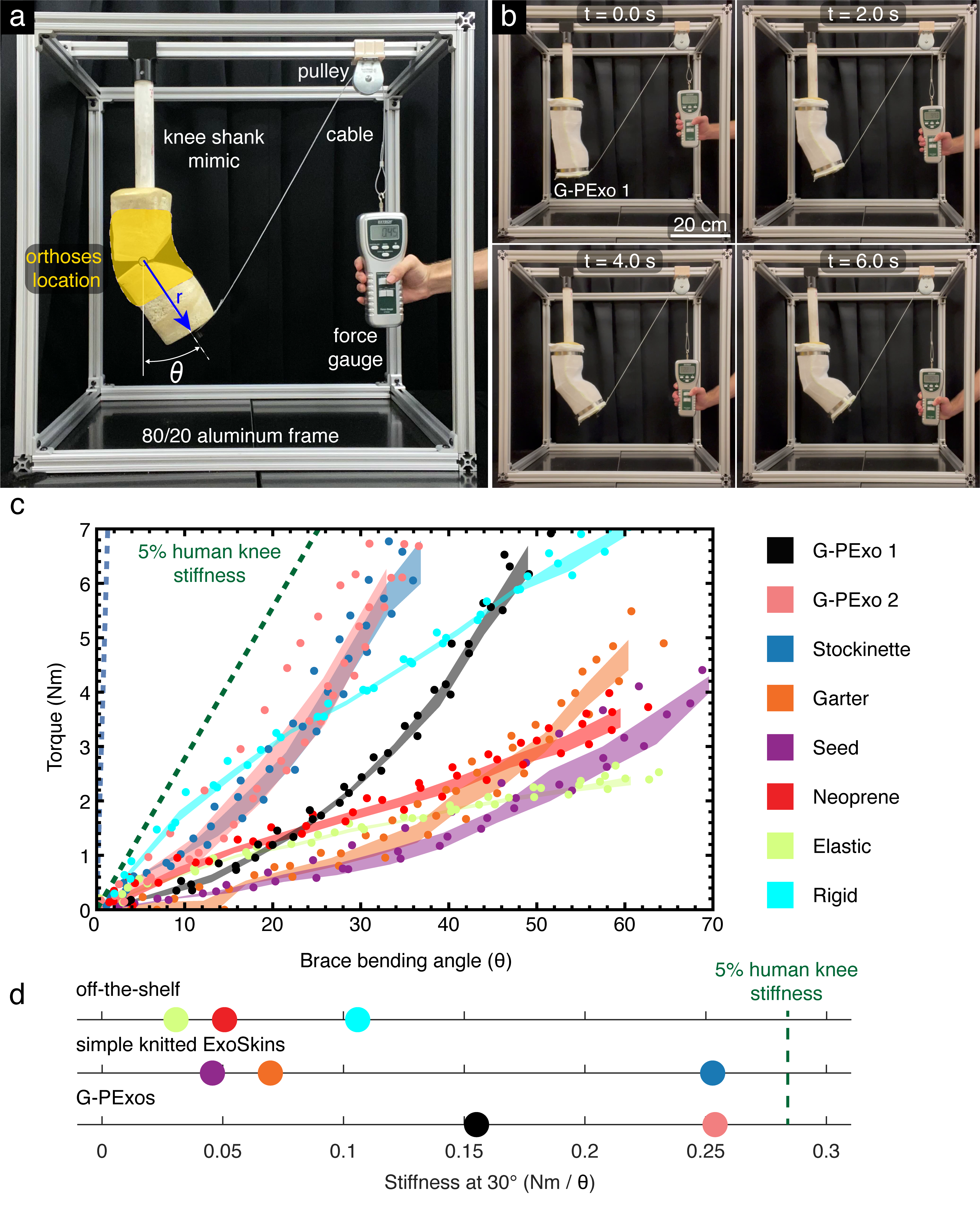}
    \caption{\textbf{Torque-rotation measurements of off-the-shelf orthoses, simple knitted ExoSkins, and G-PExos.} An in-house test rig shown in (\textbf{a}) was designed to measure the torque-angle response of an orthosis undergoing knee flexion as shown in (\textbf{b}). Torque versus angle measurements extracted from experiments on all orthoses used in this study are shown in (\textbf{c}), where solid points represent the raw experimental torque versus knee angle data and the width of the translucent regions is one standard deviation of the three experiments performed for each orthosis. The dashed blue line represents the stiffness of the human knee and the dashed forest green line represents 5\% stiffness\cite{shamaei_biomechanical_2015}. In (\textbf{d}), we plot the stiffness at $\theta = 30^{\circ}$ for each orthosis.}
    \label{fig:fig3}
\end{figure}

\begin{figure}
    \centering
    \includegraphics[width = 1\textwidth]{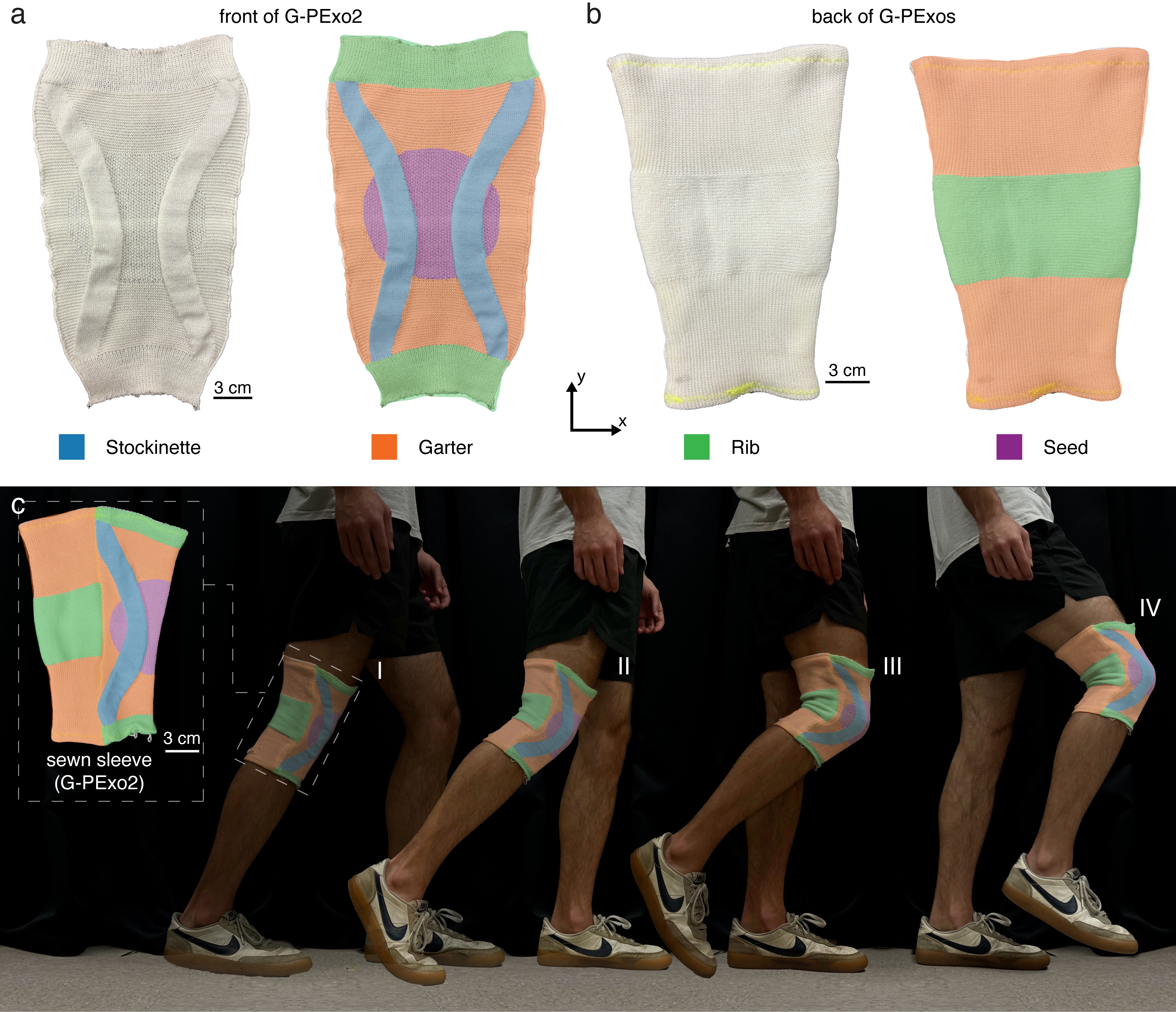}
    \caption{\textbf{Geometrically programmed Exoskin (G-PExo) designs.} Four types of knitted fabrics are strategically placed on the (\textbf{a}) front and (\textbf{b}) back of the G-PExos to achieve desired response. An image of a user wearing G-PExo2 during different phases of walking (I-IV) is shown in (\textbf{c}).}
    \label{fig:wearingsleeve}
\end{figure}

\begin{figure}
    \centering
    \includegraphics[width = 1\textwidth]{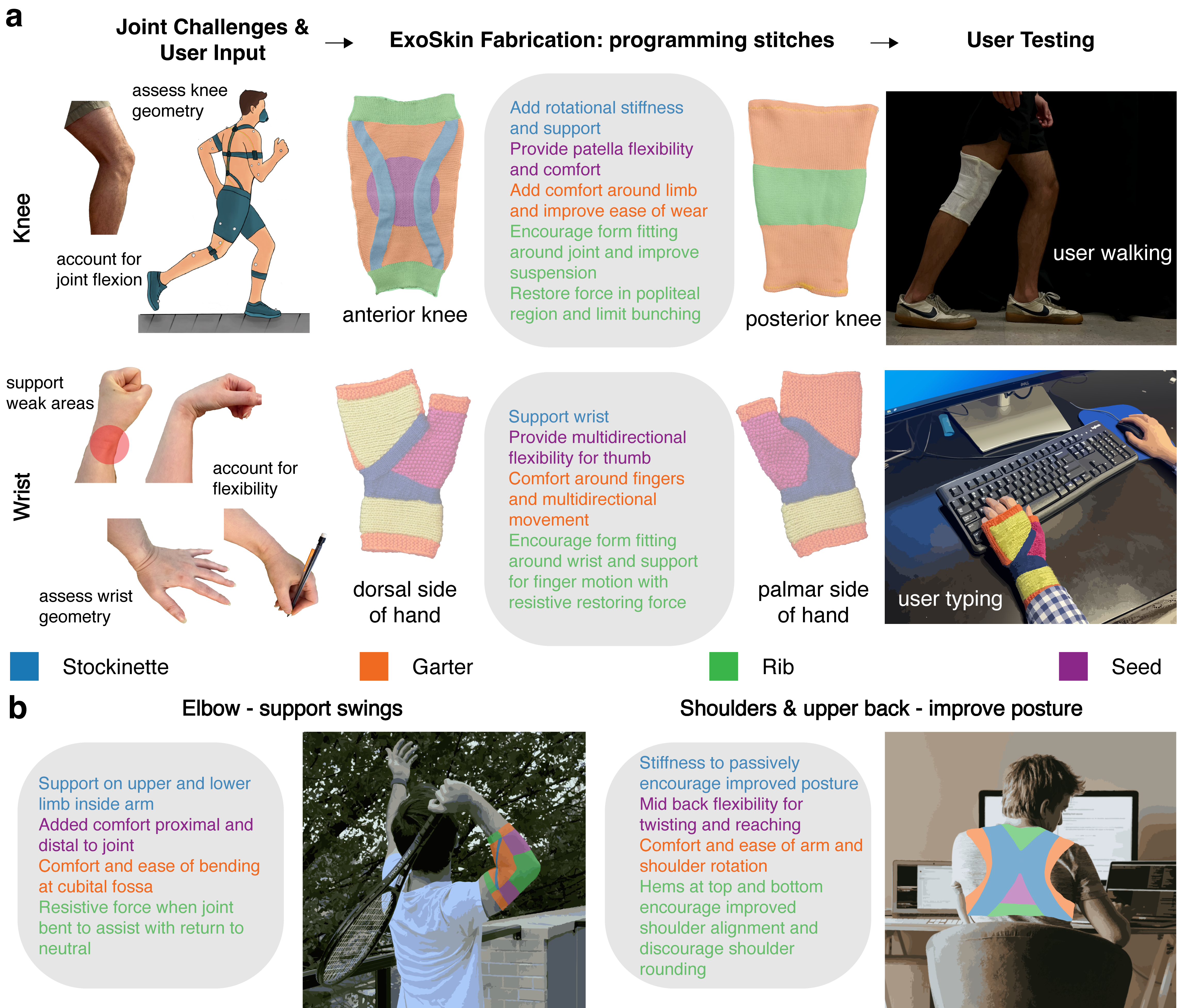}
    \caption{\textbf{Current and conceptual ExoSkins for different parts of the body.} (\textbf{a}) Workflow of ExoSkin development starting from identification of joint challenges and user specific geometry to tests ``in the wild.'' We focus on two examples including the knee discussed in this study as well as the wrist shown as a potential application in Singal \textit{et al.}\cite{singal_programming_2024}. (\textbf{b}) Future potential ExoSkin designs and applications for both the elbow and shoulders/upper back. Our proposed flow involves identifying the challenges, the ways to address them in programmable knitted designs, and then studying its applications ranging from walking, sports, and sitting for long periods at time at a desk.}
    \label{fig:fig5}
\end{figure}

Torque angle experiments were performed on the three simple knitted ExoSkins, the two G-PExos, and the three off-the-shelf knee orthoses. Experimentation began by performing a control test without any orthosis to 1) establish a baseline for correlating internal knee angle with cable travel (Figure \ref{fig:fig3}a), since the angle measurement was obscured by the orthoses during testing; and 2) to characterize any torque angle response associated with the test rig itself, which was subtracted from the response obtained from the ExoSkins or off-the-shelf orthoses. To obtain the torque-rotation response of each orthosis, they were placed on the foam knee replica in the same way a human user would wear them. The orthoses were fastened to the foam knee replica with clamps at the bottom of the lower limb and the top of the upper limb to prevent loss of suspension during testing. Force was steadily applied downward on the force meter while video was recorded of the rotation and force readings. The test was concluded when the knee replica reached the full range of motion (approximately $70^{\circ}$), or in some cases, when high shear forces induced slipping of the orthosis. From these videos, 15 evenly spaced frames, starting at the first frame, were extracted and imported into MATLAB. A MATLAB measuring tool\cite{neggers_measure_2023} was used to record precise lengths and angles of each frame by first calibrating a conversion from pixels to distance. With the force, $F$, angle, $\theta$, and moment arm, $r$, measurements obtained from the videos, a MATLAB script was used to calculate $\tau = F \cdot \cos(\theta) \cdot r$ for each of the fifteen knee angles. A total of 45 torque-angle data points are plotted in Figure \ref{fig:fig3}c for each orthosis (15 points from each of three repetitions of the test) along with second order polynomial fits obtained using least squares regression. A similar fit for the control test (i.e., without any orthosis) was subtracted from those for each orthosis to isolate the effects of the orthosis before plotting the final fitted curves in Figure \ref{fig:fig3}c. The stiffness at $30^{\circ}$ was taken as the first derivative of each fitted curve and plotted in Figure \ref{fig:fig3}d.

\clearpage

\begin{table}[ht]
\centering
\begin{tabular}{|c|c|c|c|c|}
\hline
Swatch & $Y_x$ (N/mm) & $R^2$ & $Y_y$ (N/mm) & $R^2$ \\
\hline\hline
Neoprene & 0.757 & 0.999 & 1.009 & 0.999 \\
\hline
Elastic & 0.500 & 0.992 & 0.376 & 0.991 \\
\hline
Stockinette & 0.160 & 0.915 & 0.942 & 0.980 \\
\hline
Garter & 0.114 & 0.890 & 0.147 & 0.981 \\
\hline
Rib & 0.020 & 0.943 & 0.271 & 0.975 \\
\hline
Seed & 0.072 & 0.927 & 0.167 & 0.982 \\
\hline
\end{tabular}
\caption{\label{tab:rigidity2} Young's Moduli in the $x$- and $y$-direction,  $Y_x$ and $Y_y$, for each of the swatches that underwent uniaxial stretching experiments. 
}
\end{table}

\clearpage

\clearpage

\clearpage

\clearpage

\clearpage

\clearpage

\section*{Acknowledgments}

E.D.S, G.S, and E.A.M acknowledge support from the Woodruff Launch Seed Grant from the George W. Woodruff School of Mechanical Engineering and the Building Teams Seed Grant from the College of Engineering at Georgia Tech. E.D.S and E.A.M acknowledge support from the IBB Interdisciplinary Research Seed Grant from the Institute for Bioengineering and Biosciences at Georgia Tech. A.K.S. thanks the Alexander von Humboldt Foundation and the Max Planck Society for their support. 

\section*{Author contributions statement}

K.S., A.K.S., E.D.S., G.S., and E.A.M. conceived the idea. K.S. designed and fabricated all knitted orthoses and samples for uniaxial stretching experiments and torque angle experiments. K.S. designed and conducted the uniaxial stretching experiments. S.P.K. constructed the test-rig setup and performed all of the torque angle experiments. K.S. and S.P.K. analyzed the data. A.K.S., E.D.S., G.S., K.R.H. and E.A.M. supervised the research project. All authors contributed to the writing and editing of the manuscript.

\newpage

\section*{Supplementary Methods}
\renewcommand{\thefigure}{S\arabic{figure}}

\setcounter{figure}{0}

\subsection*{Uniaxial Stretching Experiments versus Torsional Stiffness Experiments}

We compared the uniaxial stretching results (\ref{fig:Efits}) of the sample swatches with the results from the torsional stiffness experiments of the simple knitted ExoSkins and two of the off-the-shelf orthoses, neoprene and elastic. 
Comparisons are shown in \ref{fig:force_angle} where the stress versus strain data is converted to force versus angle (to match the data format from the test rig). Neither the $x$-direction (\ref{fig:force_angle}a) nor the $y$-direction (\ref{fig:force_angle}b) swatch-level data match the test rig data (\ref{fig:force_angle}c). This emphasizes the importance of our torsional stiffness experiment setup, which indicates that stress-states beyond uniaxial tension play an important role in dictating the mechanical response of knee orthoses and experimental protocols that are designed to mimic knee motion are critical.

\subsection*{Torsional Stiffness Experiments - Cast Fabrication}
A positive model of the knee of one of the authors (22-year-old male, 61.2 kg) was created by first curing fiberglass casting tape around the volunteer's mid-shaft of the left leg from the calf to the thigh (\ref{fig:methods_cast}a). Casts were taken at a 45° knee flexion angle and a relaxed straight leg (\ref{fig:methods_cast}b). The cast was removed from the volunteer and closed with staples and tape to be used as positive molds (\ref{fig:methods_cast}b). A 1" diameter PVC pipe was inserted into the mold of the relaxed straight leg and a two-part curable foam (Pedilen Rigid Foam 700) was poured around it. Once the foam set, the cast was removed carefully and the foam leg was set aside for use in the test-rig. For the bent leg cast, metal tubing was inserted into the center of the mold while quick-setting plaster was poured around it (\ref{fig:methods_cast}c). Once the plaster had set, the cast was carefully removed. Any artifacts of the casting process such as seam lines were sanded away and the plaster leg was set aside for use as a reference model (\ref{fig:methods_cast}d).

\subsection*{Simple ExoSkin and G-PExo Fabrication}

The knitted ExoSkins were designed in the M1Plus software then fabricated on a STOLL Industrial Knitting Machine. We have to take into account stitch size, the amount of yarn used per stitch, when designing the ExoSkins. Regarding the simple knitted Exoskins (\ref{fig:redbraces}), the stockinette and garter sleeves were made at a stitch size of 12 while the seed sleeve was made at size 11. For the different regions on the G-PExos, stockinette and garter portions again were made at a stitch size of 12 while seed and rib portions were made at size 11 (\ref{fig:multistitch_braces}a,b). This change in size is because the stitch patterns for both rib and seed involve switching between knit and purl stitches every column. On a v-bed knitting machine, this results in using both beds of the machine at once, requiring a larger quantity of yarn to reach both beds between stitches. Therefore, to combat the increased amount of yarn typically used, we decreased the stitch size for rib and seed. 

When knitting garments on a flatbed knitting machine, they must be constructed along their $y$-axis. Because we wanted to take particular advantage of the $x$-direction properties of the fabrics for the back of the G-PExos, we constructed them horizontally to later be rotated. We used stockinette fabric for regions not being used in the G-PExos yet were needed to ensure smooth fabrication on the knitting machine. 

Each knitted orthosis in this study were made as two separate pieces and then sewn together.

\clearpage
\begin{figure}
    \centering
    \includegraphics[width = 1\textwidth]{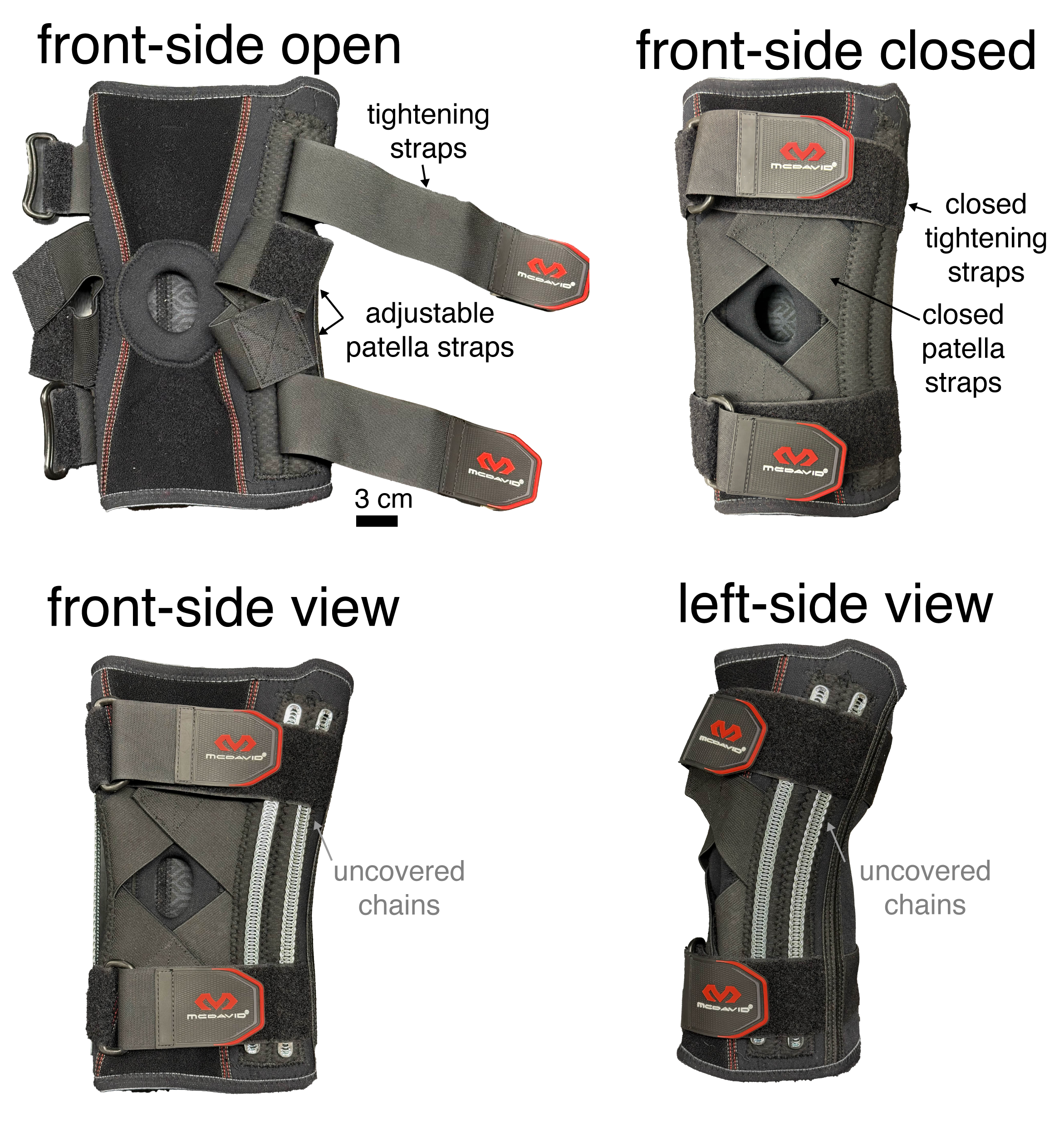}
    \caption{Anatomy of the disassembled rigid off-the-shelf orthosis shown from different angles. The McDavid 425 Ligament Knee Support with Stays and Cross Straps has several straps that tighten the orthosis on the knee and helps keep it in place. We dissect the orthosis and uncover rigid chains on the sides of the orthosis.}
    \label{fig:hardbraces}
\end{figure}

\clearpage
\begin{figure}
    \centering
    \includegraphics[width = 1\textwidth]{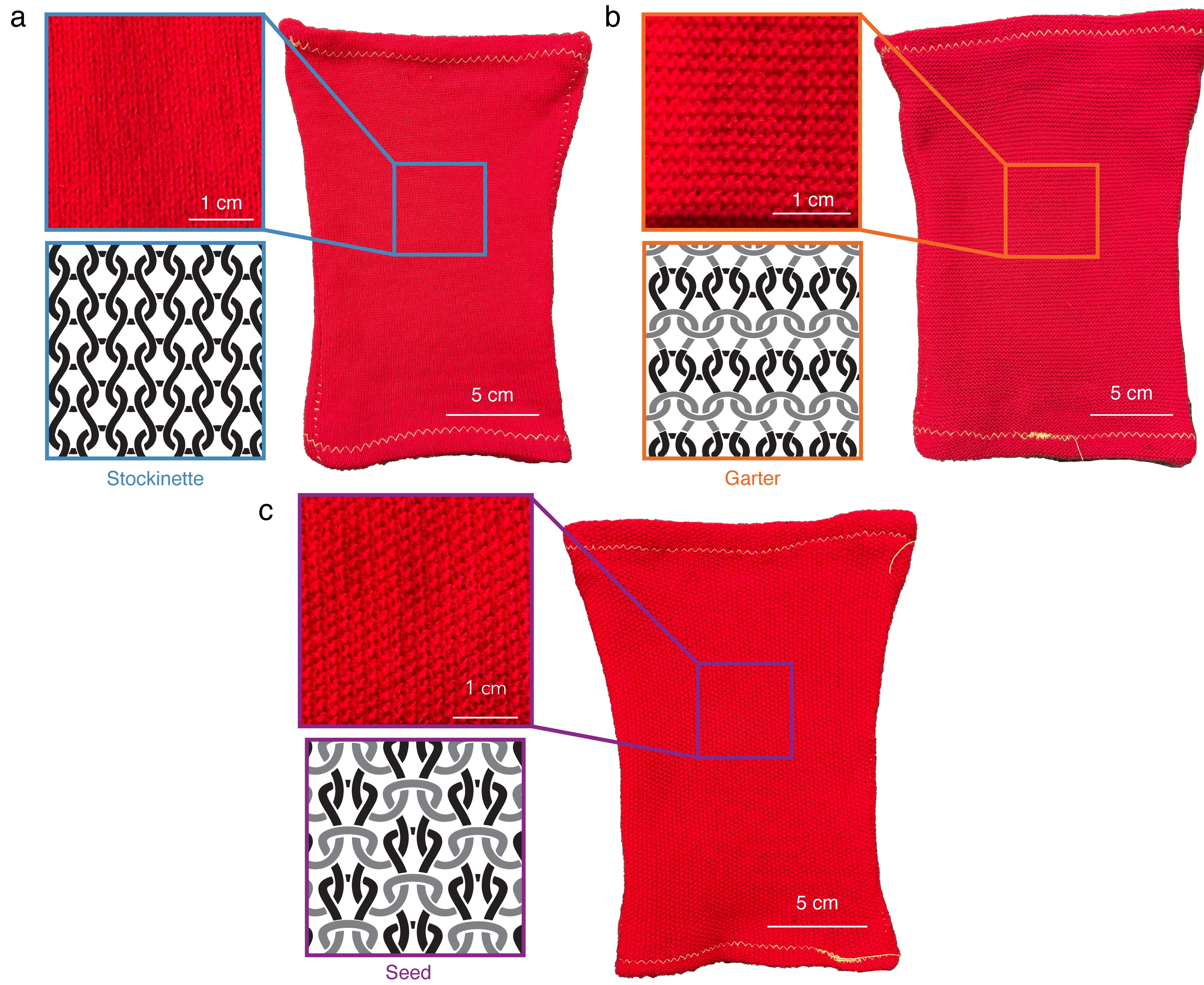}
    \caption{Simple knitted ExoSkins made of one type of knitted fabric each, (\textbf{a}) stockinette, (\textbf{b}) garter, and (\textbf{c}) seed.}
    \label{fig:redbraces}
\end{figure}

\clearpage

\begin{figure}
    \centering
    \includegraphics[width = 1\textwidth]{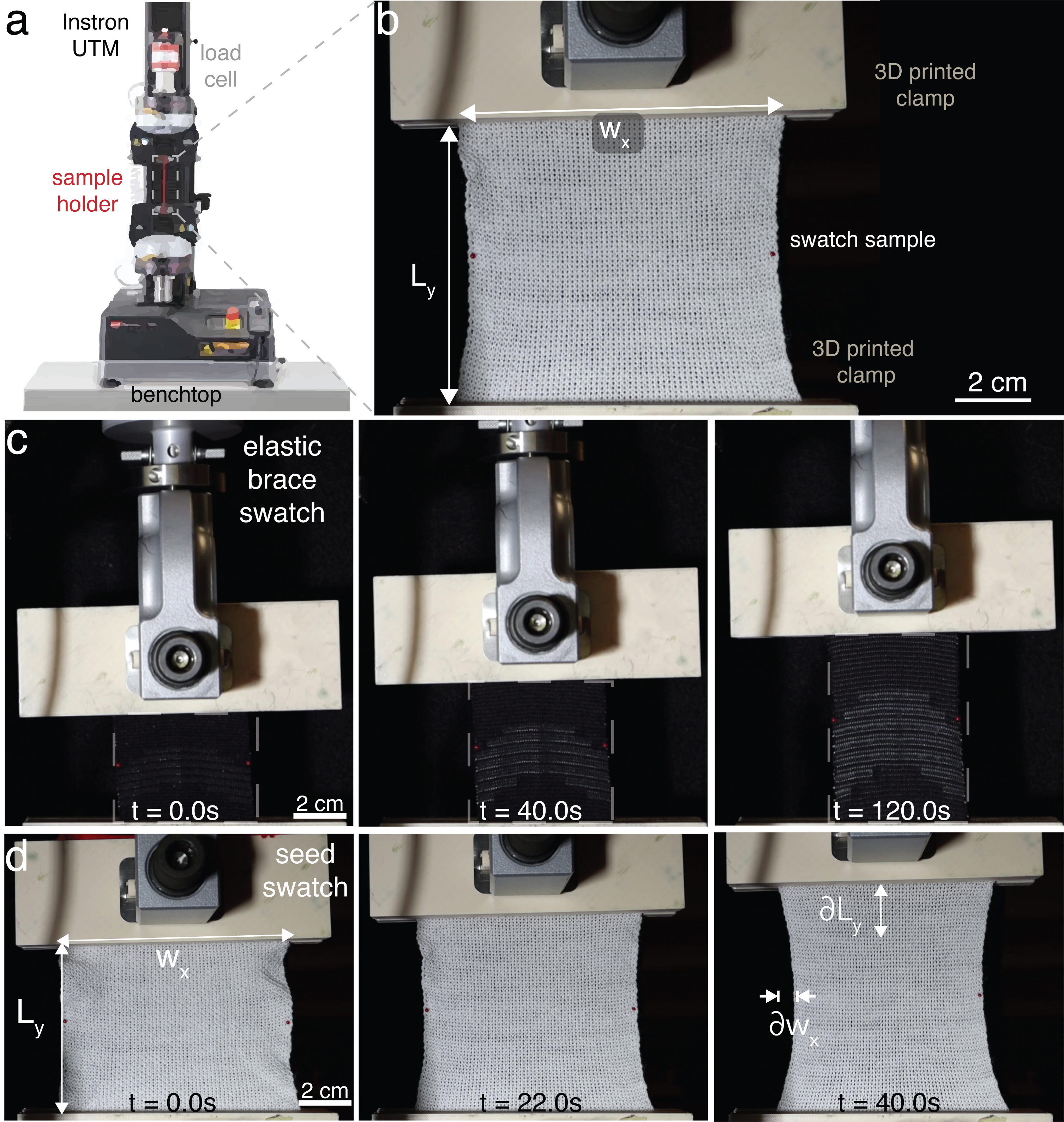}
    \caption{Uniaxial stretching experiments performed on a Universal Testing Machine (UTM). (\textbf{a}) Instron Universal Testing Machine (UTM) Model 68SC-1 with load cell and sample holder with zoom in shown in (\textbf{b}). The sample holder has custom 3D-printed clamps with saw teeth fixed to the swatch sample. Examples of experiments performed along the $y$-axis done on: (\textbf{c}) the elastic brace sample, and (\textbf{d}) the seed fabric sample.}
    \label{fig:UTM}
\end{figure}

\clearpage

\begin{figure}
    \centering
    \includegraphics[width = 1\textwidth]{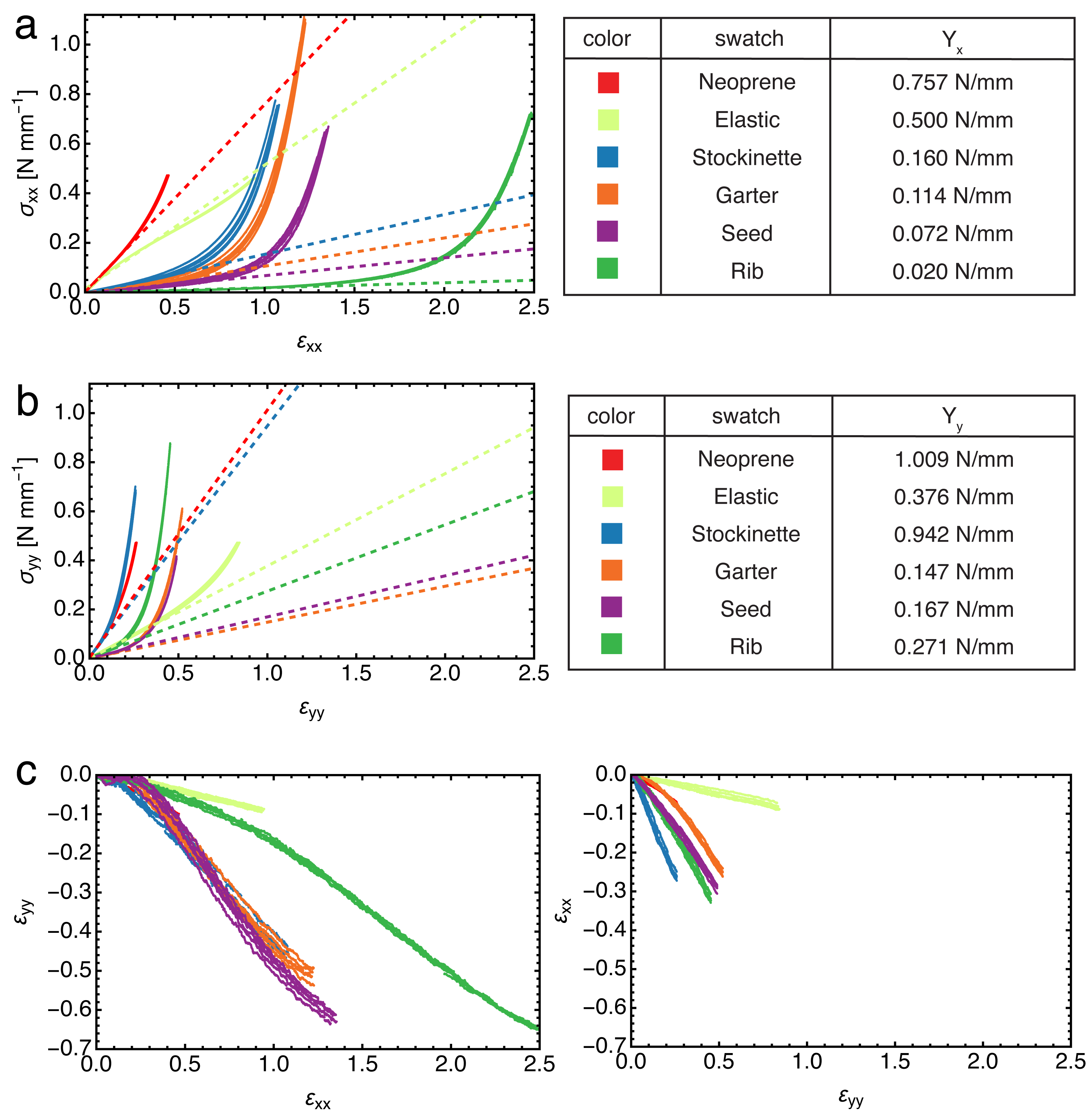}
    \caption{Stress versus strain results in the (\textbf{a}) $x$-axis and (\textbf{b}) $y$-axis for the six tested swatches from the uniaxial stretching experiments. The dotted lines represent linear fits to the low strain data and slopes of these fits represent the materials' Young's modulus. (\textbf{c}) Axial versus transverse strain plots of each of the samples stretched along either their $x$-axis (left) or their $y$-axis (right).}
    \label{fig:Efits}
\end{figure}

\clearpage

\begin{figure}
    \centering
    \includegraphics[width = 1\textwidth]{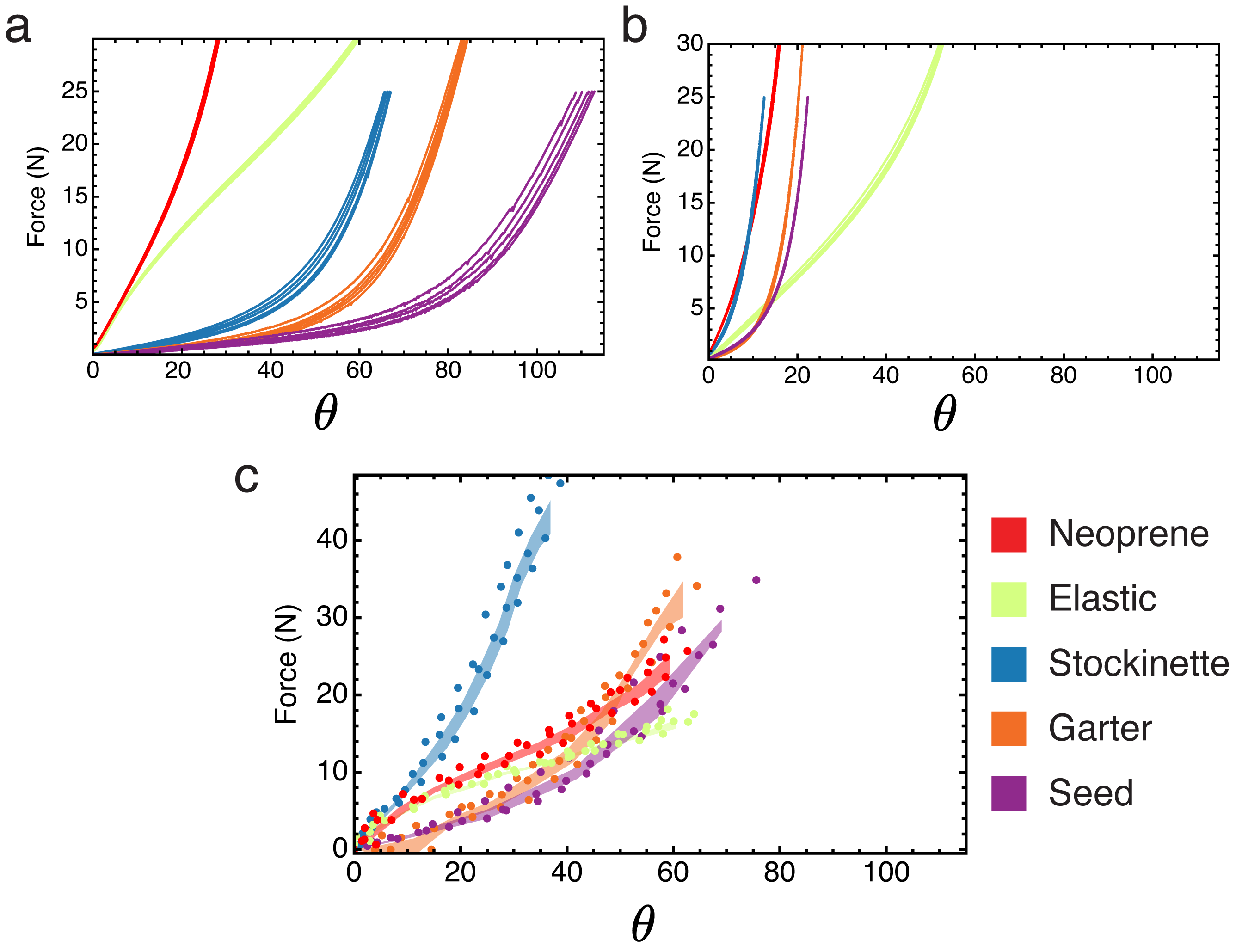}
    \caption{Force-angle comparisons between uniaxial tests performed on swatches taken from the orthoses and the rotational tests performed on the off-the-shelf orthoses and simple knitted ExoSkins. The $x$-direction (\textbf{a}) and $y$-direction (\textbf{b}) responses of the swatches do not directly match the results from the test rig (\textbf{c}).}
    \label{fig:force_angle}
\end{figure}

\begin{figure}
    \centering
    \includegraphics[width = 1\textwidth]{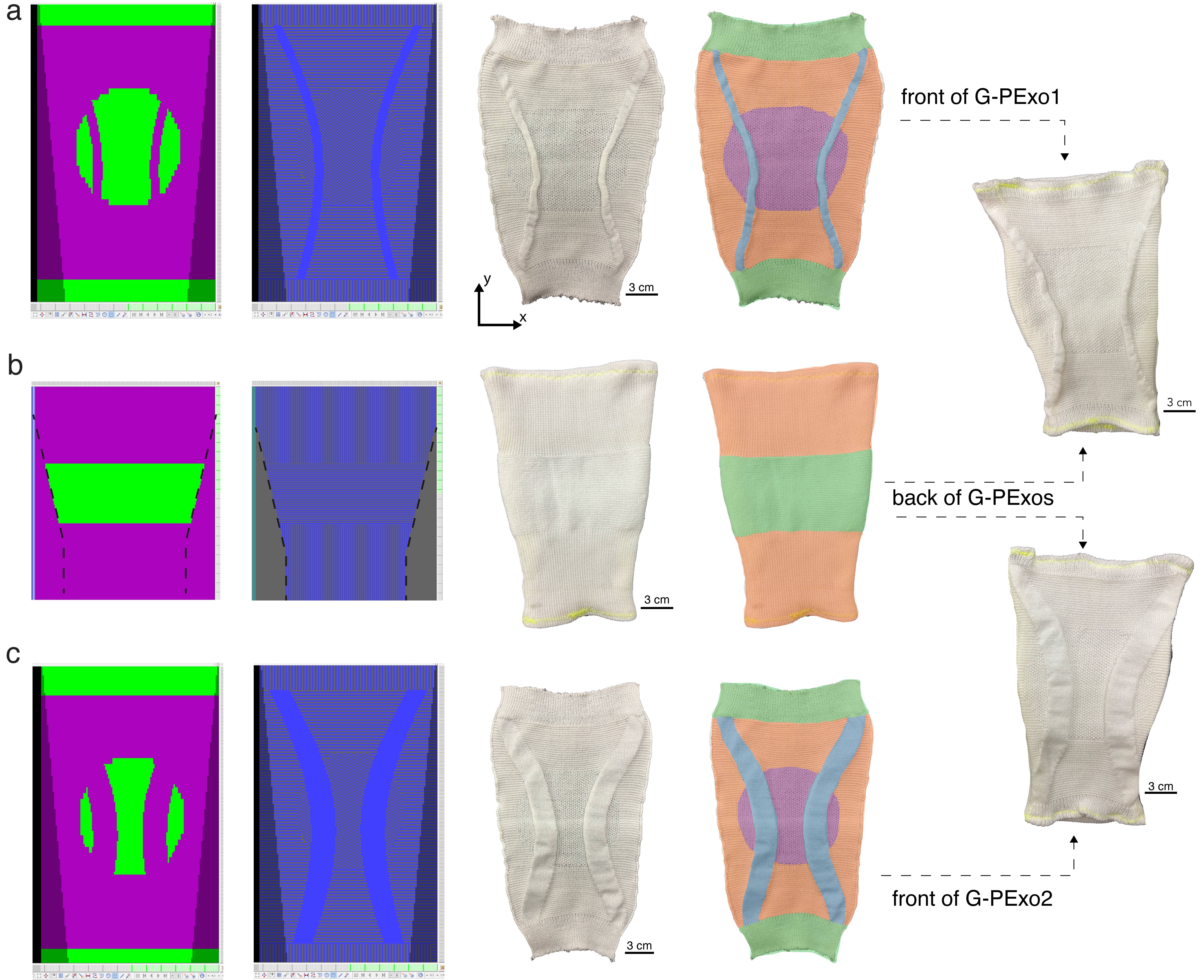}
    \caption{Intuition-based design for geometrically-programmed ExoSkins (G-PExo). We show the steps of designing the front of G-PExo1 (\textbf{a}), the back of the G-PExos (\textbf{b}), and the front of G-PExo2 (\textbf{c}). We design both G-PExos using the M1plus software (\textbf{a-c} left) and assign specific stitch sizes to the different patterns. The leftmost panel shows the stitch sizes assigned to the regions of the design (where purple is stitch size 12 and green is stitch size 11), while the next panel shows where the knit and purl stitches are assigned (blue being knit stitches and gray being purl stitches). We then show the fabricated samples made on the STOLL Industrial Knitting Machine and a color-coded overlay explicitly showing where the four types of fabrics are placed: stockinette (blue), garter (orange), rib (green), and seed (purple). The front and back panels are then sewn together to form the G-PExos (far right).}
    \label{fig:multistitch_braces}
\end{figure}

\begin{figure}
    \centering
    \includegraphics[width = 1\textwidth]{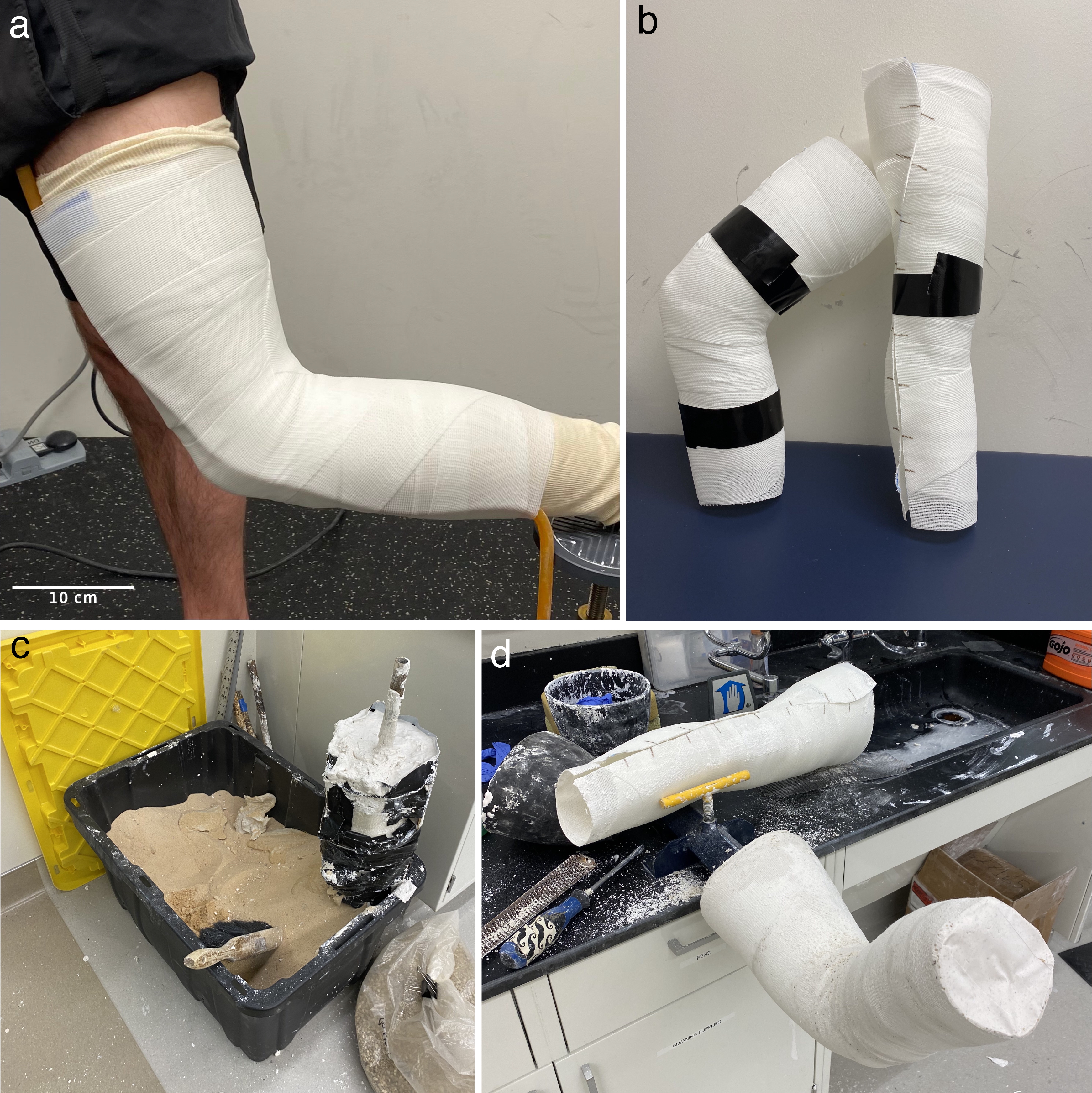}
    \caption{Knee replica fabrication for the test rig. 
    We begin by (\textbf{a}) making a cast of a human knee, (\textbf{b}) removing the cast, (\textbf{c}) filling the cast with plaster around a tubular core, and (\textbf{d}) removing the plaster mold from the cast and smoothing any irregularities. The plaster mold of the relaxed mold in (\textbf{d}) is cut at the knee joint and a hinge is inserted to create the final knee replica used in the test rig.
    A similar process was performed using two parts of curable foam for the straight leg for the test rig.}
    \label{fig:methods_cast}
\end{figure}

\end{document}